\documentstyle[11pt]{article}

\begin{document}

\title{Eichler integrals and String Theory\footnote{Article published in the Journal of Mathematical Physics 37 (1996).}}

\author{Leonidas Sandoval Junior\\ King's College London\thanks{Current address: Department of Mathematics, Universidade do Estado de Santa Catarina (UDESC),  e-mail: dma2lsj@dcc.fej.udesc.br}}

\maketitle 

\begin{abstract}
In this article, it is shown how to obtain objects called {\sl Eichler integrals} in the mathematical literature that can be used for calculating scattering amplitudes in String Theory. These Eichler integrals are also new examples of Eichler integrals with poles.
\end{abstract}

\section{Introduction}

The concept of an Eichler integral is closely related to the concept of automorphic forms. Although automorphic forms have a large range of applications in Physics and, in particular in String Theory, Eichler integrals remain relatively unknown objects to both mathematicians and physicists. One can picture an Eichler integral as a generalization of the concept of automorphic forms, and they are related to the better known {\sl Beltrami differentials} that are used in String Theory, in particular in the calculation of multiloop scattering amplitudes of strings.

This article begins with a description of the main properties of automorphic forms with some examples that will be useful when we describe the new Eichler integrals. The definition of Eichler integrals is given next, with some examples that can be used in String Theory.

\section{Automorphic Forms}

An automorphic form of weight $q$ is a function \cite{1} $\phi (z)$ that transforms in the following way under a projective transformation $P_a$:
\[ \phi \left( P_a(z)\right) =\left[ \frac{\partial P_a(z)}{\partial z}\right] ^q\phi (z)\ .\] 
Automorphic forms with just one pole will be more of our interest, since the order of the pole is limited in a simple way by the Riemann-Roch theorem \cite{2}. We now give some examples of these functions.

\subsection{Example 1}

A simple case of an automorphic form of weight $q$ can be given for the case where there is just one projective transformation $P(z)$ given by
\[ P(z)=w(z-\alpha )+\alpha \ ,\] 
where $\alpha $ is the finite fixed point of the transformation (the other fixed point is at infinity) and $w$ is the multiplier. An automorphic form $\phi (z)$ of weight $q$ can then be given by
\[  \phi (z)=(z-\alpha )^q\ .\] 

In the case where $P(z)$ has two finite fixed points, when it can be expressed by
\[ P(z)=\frac{\alpha (x-\beta )-w\beta (z-\alpha )}{(z-\beta )-w(z-\alpha )}\ ,\] 
an automorphic form of weight $q$ is given by
\[ \phi (z)=\left[ (z-\alpha )(z-\beta )\right] ^q\ .\] 

\subsection{Example 2}

In the case of a single projective transformation $P(z)$ with just one fixed point, we may have automorphic forms with poles, given by
\[ \phi _n(z)=(z-\alpha )^q\left( \frac{\partial }{\partial t}\right) ^{n-1}{\cal P}(t)\ ,\ \ t=\ln (z-\alpha )-\ln (-\alpha )\ ,\] 
where $n$ is the order of the pole and ${\cal P}(t)$ is the Wierstrass ${\cal P}$ function \cite{3} with periods $\ln w$ and $2\pi i$.

In the case of $P(z)$ having two finite fixed ponits, we have
\[ \phi (z)=\left[ (z-\alpha )(z-\beta )\right] ^q\left( \frac{\partial }{\partial t}\right) ^{n-1}{\cal P}(t)\ ,\ \ t=\ln \left( \frac{z-\alpha }{z-\beta }\right) -\ln \left( \frac{\alpha }{\beta }\right) \ ,\] 
instead.

\subsection{Example 3}

Still another way of obtaining an automorphic form on a genus $g$ Riemann surface is considering the following function \cite{1} (called a Poincar\'e series):
\[ \phi (z,\zeta )=\sum_a\left[ \frac{dT_a(z)}{dz}\right] ^{-q}\frac{\zeta -k}{T_a(z)-\zeta }\ ,\] 
where $k$ is an arbitrary constant and the sum $\sum_a$ is over all elements of the Schottky group, which is the group of all possible combinations of the projective transformations $P_b(z)$, $b=1,\dots ,g$.

Under a transformation $z\rightarrow T_b(z)$, one gets
\[ \phi \left( T_b(z),\zeta \right) =\sum_a\left[ \frac{dT_aT_b(z)}{dz}\right] ^{-q}\frac{\zeta -k}{T_aT_b(z)-\zeta }\ .\] 
We now perform a change of variables $T_aT_b(z)=T_c(z)$. Using the chain rule we have
\[ \frac{d}{dz}T_c(z)=\left[ \frac{d}{dz}T_aT_b(z)\right] \times \frac{d}{dz}T_b(z)\ ,\] 
so that we have
\[ \phi \left( T_b(z),\zeta \right) =\left[ T'_b(z)\right] ^q\sum_a\left[ \frac{dT_c(z)}{dz}\right] ^{-q}\frac{\zeta -k}{T_c(z)-\zeta }\ .\] 
Since we are summing over all the elements of the Schottky group, the series above is equivalent to the original one, so that we have
\[ \phi \left( T_b(z),\zeta \right) =\left[ T'_b(z)\right] ^q\phi (z,\zeta )\ ,\] 
i.e. it is an automorphic form of weight $q$.

\subsection{Example 4}

In the multiloop case, let us consider only projective transformations $P_a$ with finite fixed points $\alpha _a$ and $\beta _a$. We then consider the following series:
\begin{equation}
\label{eq1}
P_w(z)=\sum_b\frac{(z-\alpha _b)(z-\beta _b)}{w_b(\alpha _b-\beta _b)}\frac{\delta w_b}{\epsilon }\ ,
\end{equation}
where the sum is over all the elements of the Schottky group formed by these transformations and $\delta w_b$ and $\epsilon $ are infinitesimals. We then have that
\begin{eqnarray*}
 & & P_w\left( T_a(z)\right)  =\sum_b\left[ T_a(z)-\alpha _b\right] \left[ T_a(z)-\beta _b\right] \\ 
 & &  =\frac{T'_a(z)}{w_a(\alpha _a-\beta _a)^2}\sum_b\frac{1}{w_b(\alpha _b-\beta _b)}(\alpha _a-w_a\beta _a -\alpha _b+w_a\alpha _b)(\alpha _a-w_a\beta _a-\beta _b+w_a\beta _b)\\ 
 & & \times \left[ z-\frac{(\alpha _a\beta _a-w_a\alpha _a\beta _a-\alpha _b\beta _a+w_a\alpha _b\alpha _a)}{(\alpha _a-w_a\beta _a-\alpha _b+w_a\alpha _b)}\right] \left[ z-\frac{(\alpha _a\beta _a-w_a\alpha _a\beta _a-\beta _b\beta _a+w_a\beta _b\alpha _a)}{(\alpha _a-w_a\beta _a-\beta _b+w_a\beta _b)}\right] \frac{\delta w_b}{\epsilon }\ .
\end{eqnarray*}

Making then the following change of variables:
\begin{eqnarray}
\label{eq2}
 & & \alpha _c= \frac{(1-w_a)\alpha _a\beta _a -(\beta _a-w_a\alpha _a) \alpha _b}{\alpha _a -w_a\beta _a -(1-w_a)\alpha _b}\ ,\\ 
\label{eq3}
 & & \beta _c= \frac{(1-w_a)\alpha _a\beta _a -(\beta _a-w_a\alpha _a) \beta _b}{\alpha _a -w_a\beta _a -(1-w_a)\beta _b}\ ,\\ 
\label{eq4}
 & & w_c=w_b\ ,
\end{eqnarray}
we may write
\[ P_w\left( T_a(z)\right) =T'_a(z)\sum_c\frac{(z-\alpha _c)(z-\beta _c)}{w_c(\alpha _c-\beta _c)}\frac{\delta w_c}{\epsilon }\ ,\] 
where we have used
\[ \delta w_c=\frac{\partial w_c}{\partial w_b}\delta w_b\ .\] 
Since we are summing over all elements of the Schottky group (with the condition that the fixed points are finite), we then see that the expression on the right hand side is equivalent to the series we started with, so that
\[ P_w\left( t_a(z)\right) =T'_a(z)P_w(z)\ .\] 

We may also consider the following two series:
\begin{eqnarray}
\label{eq5}
 & & P_\alpha (z)= \sum_b \frac{(1-w_b)}{w_b(\alpha _b -\beta _b)^2} (z-\beta _b)^2 \frac{\delta \alpha _b}{\epsilon }\ ,\\ 
\label{eq6}
 & & P_\beta (z)= \sum_b \frac{(1-w_b)}{w_b(\alpha _b -\beta _b)^2} (z-\alpha _b)^2 \frac{\delta \beta _b}{\epsilon }\ .
\end{eqnarray}
Using the same change of variables (\ref{eq2})-(\ref{eq4}) and the fact that
\begin{eqnarray*}
 & & \delta \alpha _c = \frac{\partial \alpha _c}{\partial \alpha _b}\delta \alpha _b =\frac{w_a(\alpha _a-\beta _a)^2}{\left[ \alpha _a-w_a\beta _a -(1-w_a)\alpha _b\right] ^2} \delta \alpha _b\ ,\\ 
 & & \delta \beta _c = \frac{\partial \beta _c}{\partial \beta _b}\delta \beta _b =\frac{w_a(\alpha _a-\beta _a)^2}{\left[ \alpha _a-w_a\beta _a -(1-w_a)\beta _b\right] ^2} \delta \beta _b\ ,
\end{eqnarray*}
we then obtain
\begin{eqnarray*}
 & & P_\alpha \left( T_a(z)\right) =T'_a(z)P_\alpha (z)\ ,\\ 
 & & P_\beta \left( T_a(z)\right) =T'_a(z)P_\beta (z)\ .
\end{eqnarray*}
We have then obtained three examples of automorphic forms with weight $1$.

\section{Eichler Integrals}

An Eichler integral of order $q$ is defined in the following way \cite{4} \cite{5}: it is a function $f(z)$ that transforms like
\[ f\left( T_a(z)\right) =\left[ \frac{\partial T_a(z)}{\partial z}\right] ^q \left[ f(z)+P^{q+1}(z)\right] \ ,\] 
where $P^{q+1}(z)$ is a polynomial at most of order $q+1$. It can be pictured as a generalization of the concept of automorphic form. Eichler integrals are related with automorphic forms in the following way \cite{4}: given an Eichler integral $f(z)$ of order $q$, we then have
\begin{equation}
\label{eq7}
\phi (z)=\left( \frac{\partial }{\partial z}\right) ^{2q+1}f(z)\ ,
\end{equation}
where $\phi (z)$ is an automorphic form of weight $q+1$, i.e.
\[ \phi \left( T_a(z)\right) =\left[ \frac{\partial T_a(z)}{\partial z}\right] ^{q+1}\phi (z)\ .\] 

Here are some examples of Eichler integrals.

\subsection{Example 1}

A trivial example of an Eichler integral is given by any polynomial of order $2$, i.e. any function of the form
\[ f(z)=a+bz+cz^2\ ,\] 
where $a$, $b$, and $c$ are constants. Such a function transforms like
\[ f\left( T_a(z)\right) =T'_a(z)\left[ f(z)+\frac{1}{w(\alpha -\beta )^2}(d+ez+fz^2)\right] \ ,\] 
where
\begin{eqnarray*}
 & & d=(1-w)^2a+(1-w)(\alpha -w\beta )b -\left[ w\beta ^2(1-w) -2w(\alpha \beta -1) +w\alpha ^2\right] c\ ,\\ 
 & & e=2(w\alpha -\beta )(1-w)a -2\alpha \beta (1-w)^2b +\left[ 2w\beta (\alpha +\beta )-2\alpha \beta (w\beta +\alpha )\right] c\ ,\\ 
 & & f=(\beta ^2-w\alpha ^2)(1-w)a +\alpha \beta (1-w)(\beta -w\alpha )b -w\alpha \beta ^2(2-w\alpha )c\ .
\end{eqnarray*}

Differentiating it three times, we obtain
\[ \frac{d^3}{dz^3}f(z)=0\ ,\] 
which is a sort of automorphic form (a trivial one) so equation (\ref{eq7}) holds.

\subsection{Example 2}

There are not many examples of Eichler integrals with poles in the literature. The first one was given by Ahlfors \cite{4}. In our notation, his Eichler integral is given by
\begin{equation}
f(z,\zeta)=\sum_b\left[ z-T_b(\zeta )\right] ^{-1} \left[ T'_b(\zeta )\right] ^q\ ,
\end{equation}
where the sum is over all elements of the Schottky group.

This function transforms in the following way:
\begin{equation}
f\left( T_a(z),\zeta\right) =\left[ T'_a(z)\right] ^{q-1}\left\{ f(z,\zeta )+\sum_b\left[ \frac{T'_aT_b(\zeta )}{\left( T'_a(z)\right) ^{q-1/2}}-1\right] \left[ z-T_b(\zeta )\right] ^{-1}\left[ T'_b(\zeta )\right] ^q\right\} \ .
\end{equation}
The second term in the right hand side can be shown to be a polynomial of degree $q$. This Eichler integral has just one simple pole at $\zeta =z$.

In order to obtain Eichler integrals with poles of higher order, one just has to form derivatives with respect to $\zeta $:
\[ f_{k+1}(z,\zeta )=\frac{\partial ^k}{\partial \zeta ^k}f(z,\zeta )\] 
where $f_{k+1}(z,\zeta )$ is an Eichler integral with a pole of order $k+1$ at $\zeta =z$.

\subsection{Example 3}

As it will be shown now, not all Eichler integrals must transform in such a complicated way. It is not hard to find examples of Eichler integrals with simple and useful transformations.

As explained before, an Eichler integral can be obtained by simply integrating an automorphic form a certain number of times. It is possible to show that every automorphic form of weight $q$ can be expressed in terms of the following Poincar\'e series:
\begin{equation}
\label{eq8}
g(z,\zeta)=\sum_b\left[ T'_b(z)\right] ^{-q}\frac{\zeta-k}{T_b(z)-\zeta}\ ,
\end{equation}
where $k$ is an arbitrary constant.

If we now take the case $q=-1$ we have simply
\[ g(z,\zeta) =\sum_b\left[ T'_b(z)\right] ^{-1}\frac{\zeta-k}{T_b(z)-\zeta} =\sum_b(\zeta -k)\left[ \frac{1}{z+(b_b-d_d\zeta)/(a_b-c_b\zeta)}-\frac{1}{z+d_b/c_b}\right] \ ,\] 
and, by (\ref{eq7}), we may obtain an Eichler integral of weight zero simply by integrating this expression once. Doing this we obtain the following:
\begin{equation}
f(z,\zeta )=\sum_b(\zeta -k)\ln \left( \frac{c_b}{(a_b-c_b\zeta )}\frac{\left[ (a_b-c_b\zeta )z+(b_b-d_b\zeta )\right] }{(c_bz+d_b)}\right) +c'\ ,
\end{equation}
where $c'$ is a constant of integration that can be set to zero.

We may then make the following change of coefficients:
\begin{eqnarray}
a_m=a_b-c_b\zeta\ \ & , &\ \ c_m=c_b\ ,\nonumber \\ 
\label{eq9}
b_m=b_b-d_b\zeta\ \ & , &\ \ d_m=d_b\ ,
\end{eqnarray}
and the function $f(z,\zeta )$ becomes simply
\[ f(z,\zeta)=\sum_m(\zeta-k)\ln \left( \frac{c_m (a_mz+b_m)}{a_m (c_m+d_m)}\right) \ .\] 

Performing the conformal transformation
\begin{equation} 
T_a(z)=\frac{a_az+b_a}{c_az+d_a}
\end{equation}
with $c_a\neq 0$ in the expression above, we obtain
\begin{equation}
\label{eq10}
f\left( T_a(z),\zeta\right) =\sum_m(\zeta-k)\ln \left( \frac{c_m \left[ a_m(a_az+b_a)+b_m(c_az+d_a)\right] }{a_m \left[ c_m(a_az+b_a)+d_m(c_az+d_a)\right] }\right) .
\end{equation}

If we now make another change of variables,
\begin{eqnarray}
a_b=a_ma_a+b_mc_a\ \ & , &\ \ c_b=c_ma_a+d_mc_a\ ,\nonumber \\ 
\label{eq11}
b_b=a_mb_a+b_md_a\ \ & , &\ \ d_b=c_mb_a+d_md_a\ ,
\end{eqnarray}
we then have
\begin{eqnarray}
f\left( T_a(z),\zeta\right)  & = & \sum_b(\zeta -k)\ln \left( \frac{(c_bd_a-c_ad_b)}{(a_bd_a-c_ab_b)}\frac{(a_bz+b_b)}{(c_bz+d_b)}\right) \nonumber \\ 
\label{eq12}
 & = & \sum_b(\zeta -k) \left[ \ln \left( \frac{c_b (a_bz+b_b)}{a_b (c_bz+d_b)} \right) +\ln \left( \frac{a_b (c_bd_a-c_ab_b)}{c_b (a_bd_a-c_ab_b)}\right) \right] \ .
\end{eqnarray}

Since we are summing over all transformations $T_b(z)$ with $c_b\neq 0$, the first term of expression (\ref{eq12}) is just $f(z,\zeta )$ and we then have
\begin{equation}
\label{eq13}
f\left( T_a(z),\zeta\right) =f(z,\zeta)+\sum_b (\zeta -k)\ln \left( \frac{a_b(d_ba_a-c_bb_a)}{c_b(b_ba_a-a_bb_a)}\right) \ .
\end{equation}
The second term of expression (\ref{eq13}) can be readily identified as a constant, i.e. a polynomial in $z$ of degree 0, so that it is proved that the function $f(z,\zeta )$ is an Eichler integral with weight 0.

We now consider automorphic form (\ref{eq8}) with weight $-2$:
\begin{equation}
g(z,\zeta)=\sum_b\left[ T'_b(z)\right] ^2\frac{\zeta-k}{T_b(z)-\zeta}\ ,
\end{equation}
where $k$ is a constant and the sum is over all elements of the Schottky group. According to the theory of Eichler integrals, if we integrate an automorphic form of weight $-2$ three times we shall obtain an Eichler integral of weight 1.

If we take $T_b(z)$ to be of the form
\[ T_b(z)=\frac{a_bz+b_b}{c_bz+d_b}\] 
we may then write this function in the following form:
\[ g(z,\zeta)=\sum_b\frac{\zeta -k}{a_b-c_b\zeta}\frac{1}{(c_bz+d_b)^3\left[ z+(b_b-d_b\zeta )/(a_b-c_b\zeta )\right] }\ .\] 
This can be easily integrated. Considering the case where $c_b\neq 0$ and performing integration three times on this automorphic form we obtain the following function:
\begin{eqnarray}
f(z,\zeta) & = & \sum_b\frac{1}{2}(\zeta-k)\left[ (a_b-c_b\zeta)z+(b_b-d_b\zeta)\right] ^2\nonumber \\ 
 & & \times \ln \left( \frac{c_b}{(a_b-c_b\zeta)}\frac{\left[ (a_b-c_b\zeta)z+(b_b-d_b\zeta)\right] }{(c_b+d_b)}\right) \nonumber \\ 
\label{eq14}
 & & +\frac{1}{2c_b(a_b-c_b\zeta)}z-\frac{2(a_b-c_b\zeta )d_b+1}{4c_b^2(a_b-c_b\zeta)^2}+c_1z^2+c_2z+c_2
\end{eqnarray}
where $c_1$, $c_2$ and $c_3$ are constants resulting from the integrations. Choosing these integration constants in such a way as to cancel with the polynomial part of expression (\ref{eq14}) we then obtain
\begin{eqnarray}
f(z,\zeta) & = & \sum_b\frac{1}{2}(\zeta-k)\left[ (a_b-c_b\zeta)z+(b_b-d_b\zeta)\right] ^2\nonumber \\ 
\label{eq15}
 & & \times \ln \left( \frac{c_b}{(a_b-c_b\zeta)}\frac{\left[ (a_b-c_b\zeta)z+(b_b-d_b\zeta)\right] }{(c_b+d_b)}\right) .
\end{eqnarray}
We can now make the change of coefficients (\ref{eq9}) and the function $f(z,\zeta )$ becomes
\begin{equation}
\label{eq16}
f(z,\zeta)=\sum_m\frac{1}{2}(\zeta-k)(a_mz+b_m)^2\ln \left( \frac{c_m(a_mz+b_m)}{a_m(c_m+d_m)}\right) .
\end{equation}

This expression should be an Eichler integral and we are going to show it indeed is. If we perform the conformal transformation
\begin{equation} 
\label{eq17}
T_a(z)=\frac{a_az+b_a}{c_az+d_a}
\end{equation}
with $c_a\neq 0$ in the expression above, we obtain
\begin{eqnarray}
f\left( T_a(z),\zeta \right)  & = & \sum_m\frac{1}{2}(\zeta-k) \left[ \frac{a_m(a_az+b_a) +b_m(c_az+d_a)}{c_az+d_a}\right] ^2\nonumber \\ 
\label{eq18}
 & & \times \ln \left( \frac{c_m\left[ a_m(a_az+b_a) +b_b(c-iz+d_a)\right] }{a_m\left[ c_m(a_az+b_a) +d_m(c_az+d_a)\right] }\right) .
\end{eqnarray}
If we now make another change of variables, given by (\ref{eq11}), we then have
\begin{eqnarray}
f\left( T_a(z),\zeta \right)  & = & T'_a(z) \sum_b\frac{1}{2}(\zeta -k)(a_bz+b_b)^2\ln \left( \frac{(d_ba_a-c_bb_a) (a_bz+b_b)}{(b_ba_a-a_bb_a)(c_bz+d_b)}\right) \nonumber \\ 
\label{eq19}
 & = & T'_a(z) \sum_b\frac{1}{2}(\zeta -k)(a_bz+b_b)^2\left[ \ln \left( \frac{c_b(a_bz+b_b)}{a_b(c_bz+d_b)}\right) +\ln \left( \frac{a_b(d_ba_a-c_bb_a)}{c_b(b_ba_a-a_bb_a)}\right) \right] \ .
\end{eqnarray}
Since we are summing over all transformations $T_b(z)$ with $c_b\neq 0$, the first term of expression (\ref{eq19}) is just $f(z,\zeta )$, and we then have
\begin{equation}
\label{eq20}
f\left( T_a(z),\zeta \right) =T'_a(z)\left[ f(z,\zeta) +\sum_b\frac{1}{2}(\zeta -k)(a_bz+b_b)^2\ln \left( \frac{a_b(d_ba_a-c_bb_a)}{c_b(b_ba_a-a_bb_a)}\right) \right] \ .
\end{equation}

The second term of expression (\ref{eq20}) can be readly identified with a polynomial in $z$ of degree 2 so that it is proved that the function $f(z,\zeta )$ is an Eichler integral.

\section{Example 4}

We consider now the case
\begin{equation}
\label{eq21}
T(z)=\frac{az+d}{cz+d}\ ,
\end{equation}
that can be written in terms of the finite fixed point $\alpha $ and the multiplier $w$ as
\begin{equation}
\label{eq22}
T(z)=w(z-\alpha )+\alpha \ .
\end{equation}

In String Theory, and in particular in the Group Theoretic approach \cite{6}, we will be interested in functions that are associated with conformal transformations that cause infinitesimal changes in the moduli of a Riemann surface with genus $g$. In one loop, i.e. a Riemann surface with genus $1$, we have one finite fixed point $\alpha $, one fixed point at infinity, and the multiplier $w$. In order to fix the modular invariance of the theory, we must fix three variables. Besides the one fixed point that has been already fixed at infinity, we can choose to fix the finite fixed point and one of the variables associated with the incoming strings so that the multiplier $w$ will be the only variable associated with the loop left.

We then want a function that has the effect of changing the multiplier $w$ infinitesimally, i.e. we want a function that transforms like
\begin{equation}
\label{eq23}
f\left( T(z)\right) =T'(z)f(z)-\frac{\partial T(z)}{\partial w}\frac{\delta w}{\epsilon }\ ,
\end{equation}
i.e.
\begin{equation}
\label{eq24}
f\left( T(z)\right) =T'(z)\left[ f(z)-\frac{\delta w}{\epsilon w}(z-\alpha )\right] 
\end{equation}
and
\begin{equation}
\label{eq25}
f\left( S(z)\right) =f(z)\ ,
\end{equation}
where $S(z)={\rm e}^{2\pi i}z$ and where $\epsilon $ and $\delta w$ are infinitesimals.

A function that transforms in this way was found in references \cite{6} and \cite{7} in the context of String Theory. It is given by
\begin{equation}
\label{eq26}
f(z)=\frac{\delta w}{\epsilon w}(z-\alpha )\bar \zeta \left( \ln (z-\alpha )\right) \ ,
\end{equation}
where the function $\bar \zeta (t)$ is given by
\[ \bar \zeta (t)=\zeta (t)-\frac{\zeta (\pi i)}{\pi i}t\ .\] 
The function $\zeta (t)$ is Weierstrass' $\zeta $ function,
\[ \zeta (t)=\frac{1}{t}+\sum_{p\neq 0}\left( \frac{1}{t-p}+\frac{1}{p}+\frac{1}{p^2}\right) \ ,\] 
where $p$ is the semi-period of the function and is given by $p=nw_1+mw_2$, ($n,m\in {\bf Z}$), where
\[ w_1=\ln w\ ,\ \ w_2=2\pi i\ .\] 

The Weierstrass $\zeta $ function transforms in the following way:
\begin{eqnarray*}
 & & \zeta (t+w_1) =\zeta (t)+2\zeta (w_1/2)\ ,\\ 
 & & \zeta (t+w_2) =\zeta (t)+2\zeta (w_2/2)\ ,
\end{eqnarray*}
and the term $\ln (z-\alpha )$ transforms like
\begin{eqnarray*}
 & & \ln \left( T(z)-\alpha \right) =\ln w+\ln (z-\alpha )\ ,\\ 
 & & \ln \left( {\rm e}^{2\pi i}z -{\rm e}^{2\pi i}\alpha \right) =\ln (2\pi i) +\ln (z-\alpha )\ ,
\end{eqnarray*}
so that we have
\begin{equation}
\label{eq27}
f\left( T(z)\right) =T'(z) \left[ f(z) -\frac{\delta w}{\epsilon w}(z-\alpha )\right] \ .
\end{equation}

The function $\bar \zeta (t)$ can be related to the theta function \cite{3} in the following way:
\[ \bar \zeta (t)=\frac{d}{dt}\ln \theta (t,\tau )\ ,\] 
where $\theta (t,\tau )$ has periods $w_1=\ln w$ and $w_2=2\pi i$.

If we take the third derivative of the function $f(z)$, we obtain
\[ g(z)=f'''(z)=\frac{\delta w}{\epsilon w}\frac{1}{(z-\alpha )^2}\left[ {\cal P}(t)+\frac{\zeta (\pi i)}{\pi i}-{\cal P}''(t)\right] \ ,\] 
where $t=\ln (z-\alpha )$ and ${\cal P}(t)$ is Weierstrass' ${\cal P}$ function with periods $w_1=\ln w$ and $w_2=2\pi i$, given by
\[ {\cal P}(t)=-\frac{d}{dt}\zeta (t)\] 
which transforms like
\[ {\cal P}(t+w_1)={\cal P}(t+w_2)={\cal P}(t)\ ,\] 
i.e. it is an elliptic function.

Since ${\cal P}(t)$ and its derivatives do not change under a transformation of $t+\ln w$ or $t+2\pi i$, we then have
\[ g\left( T(z)\right) =\frac{1}{w^2}g(z)=\left[ T'(z)\right] ^{-2}g(z)\ ,\] 
i.e. $g(z)$ is an automorphic form with weight $-2$, as expected from the relation (\ref{eq7}). So we have verified that $f(z)$ is an Eichler integral.

\subsection{Example 5}

We shall now search for a function with transformation properties similar to those of (\ref{eq23}), but now for the case of many projective transformations. We want a function that transforms like
\begin{equation}
\label{eq28}
f_w\left( T_a(z)\right) =T'_a(z)f_w(z) -\frac{\partial T_a(z)}{\partial w_a} \frac{\delta w_a}{\epsilon }\ ,
\end{equation}
where
\[ T_a(z)=\frac{\alpha _a(z-\beta _a)-w_a \beta _a(z-\alpha _a)}{(z -\beta _a)-w_a(z-\alpha _a)}\ ,\] 
for every $a=1,\dots ,g$, i.e. the action of $T_a(z)$ on this function causes infinitesimal changes in the multipliers $w_a$. So this function must transform like
\begin{equation}
\label{eq29}
f_w\left( T_a(z)\right) =T'_a(z)\left[ f_w(z) +\frac{(z -\alpha _a)(z -\beta _a)}{w_a(\alpha _a -\beta _a)} \frac{\delta w_a}{\epsilon }\right] \ .
\end{equation}
In addition to this, we also demand that
\begin{equation}
\label{eq30}
f_a\left( S_a(z)\right) =f_a(z)\ ,
\end{equation}
where $z\rightarrow S_a(z)$ is the transformation that takes $z$ once around the $a_a$ loop for $a=1,\dots ,g$.

In order to find the Eichler integral that transforms like this, we shall make analogies between the function in one loop and the function that we must have for the multiloop case. First we notice that the series $P_w(z)$ defined in (\ref{eq1}),
\[ P_w(z)= \sum_b\frac{(z-\alpha _b)(z-\beta _b)}{w_b(\alpha _b-\beta _b)}\frac{\delta w_b}{\epsilon }\ ,\] 
transforms like
\[ P_w\left( T_a(z)\right) =T'_a(z)P_w(z)\ ,\] 
so that it is the generalization for the multiloop case of the polynomial $(z-\alpha )$ for the one loop case.

Now we must try to find an analog of the Weierstrass $\zeta $ function suitable to the multiloop case. This can be obtained by first generalizing the concept of a $\theta $ function and of the Weierstrass $\zeta $ function. This function is given by the hyperelliptic $\zeta $ function \cite{8} \cite{9} or best, by the $\bar \zeta $ function defined in the Appendix. In our first attempt we attach a $\bar \zeta _b(v)$ function to every element of the series $P_w(z)$, so that we have
\[ f_{1w}(z)=\sum_b\frac{\delta w_b}{\epsilon }\frac{(z-\alpha _b)(z-\beta _b)}{w_b(\alpha _b-\beta _b)}\bar \zeta _b(v)\ .\] 
This function will not transform the way we want, since the term $(z-\alpha _b)(z-\beta _b)/(\alpha _b-\beta _b)$ and the first Abelian integrals $v_b(z)$ do not transform in the same way. We then go to the next step, which is making $\bar \zeta $ a function not of the first Abelian integrals $v_b(z)$, but of the variables $u_b(z)$ (such a change of variables is justified in Baker \cite{8}, section 192), such that
\[ u_b(z)=\ln \left( \frac{z-\alpha _b}{z-\beta _b}\right) -\ln \left( \frac{\alpha _a}{\beta _a}\right) \ .\] 
Under a change $z\rightarrow T_a(z)$, these variables will change like
\[ u_b\left( T_a(z)\right) =\frac{T_a(z)-\alpha _b}{T_a(z)-\beta _b}=\frac{(\alpha _a-w_a\beta _a-\alpha _b+w_a\alpha _b)}{(\alpha _a-w_a\beta _a-\beta _b+w_a\beta _b)} \frac{\left[ z-\frac{(\alpha _a\beta _a-w_a\alpha _a\beta _a-\alpha _b\beta _a+\alpha _bw_a\alpha _a)}{(\alpha _a-w_a\beta _a-\alpha _b+w_a\alpha _b)}\right] }{\left[ z-\frac{(\alpha _a\beta _a-w_a\alpha _a\beta _a-\beta _b\beta _a+\beta _bw_a\alpha _a)}{(\alpha _a-w_a\beta _a-\beta _b+w_a\beta _b)}\right] }\ .\] 

Performing the same change of variables as in (\ref{eq2})-(\ref{eq4}), we then obtain
\[ \frac{T_a(z)-\alpha _b}{T_a(z)-\beta _b}=w_{ca}\frac{z-\alpha _c}{z-\beta _c}\ ,\] 
where the coefficient $w_{ca}$ is given by
\[ w_{ca}=\frac{(1-w_a)\beta _c-(\beta _a-w_a\alpha _a)}{(1-w_a)\alpha _c-(\beta _a-w_a\alpha _a)}\ ,\] 
so that
\[ u_b\left( T_a(z)\right) =\ln w_{ca}+u_c(z)\ .\] 

We then redefine the generalized $\theta $ function $\theta (v)$ in the following way:
\[ \theta (u)=\sum_{n=-\infty }^\infty {\rm exp}\ \left[ \sum_{c,d=1}^g(n_c+\delta _c)\frac{1}{2}\ln w_{cd}(n_d+\delta _d)+\sum_{c=1}^g2\pi i\gamma _c(n_c+\delta _c)+\sum_{c=1}^gu_c(n_c+\delta _c)\right] \ .\] 
This function transforms like
\[ \theta (u+\Omega )={\rm exp}\ \left\{ -\sum_{c=1}^gp_c\left( u_c+\frac{1}{2}\Omega _c\right) -\sum_{c=1}^g\left[ \pi ip_cq_c-2\pi i(q_c\delta _c-p_c\gamma _c)\right] \right\} \theta (u)\ ,\] 
where $\Omega _b$ is now given by
\[ \Omega _b=\sum_{c=1}^g\left( 2\pi i\delta _{bc}p_c+\ln w_{bc}q_c\right) =2\pi ip_b+\sum_{c=1}^g\ln w_{bc}q_c\ .\] 
Defining now
\begin{equation}
\label{eq31}
\bar \zeta _b(u)=\frac{\partial }{\partial u_b}\ln \theta (u)\ ,
\end{equation}
we have
\[ \bar \zeta _b\left( u\left( T_a(z)\right) \right) =\bar \zeta _b(u)-\delta _{ab}\] 
and
\[ \bar \zeta _b\left( u\left( S_a(z)\right) \right) =\bar \zeta _b(u)\ .\] 

We then define the following function:
\begin{equation}
\label{eq32}
f(z)=\sum_b\frac{\delta w_b}{\epsilon }\frac{(z-\alpha _b)(z-\beta _b)}{w_b(\alpha _b-\beta _b)}\bar \zeta _b\left( u(z)\right) \ .
\end{equation}
This function transforms like
\begin{eqnarray*}
f_w\left( T_a(z)\right)  & = & T'_a(z)\sum_c\frac{\delta w_c}{\epsilon }\frac{(z-\alpha _c)(z-\beta _c)}{w_c(\alpha _c-\beta _c)}\left[ \bar \zeta _c\left( u(z)\right) -\delta _{ac}\right] \\ 
 & = & T'_a(z)\left[ \sum_c\frac{\delta w_c}{\epsilon }\frac{(z-\alpha _c)(z-\beta _c)}{w_c(\alpha _c-\beta _c)}\bar \zeta _c\left( u(z)\right) -\frac{\delta w_a}{\epsilon }\frac{(z-\alpha _a)(z-\beta _a)}{w_a(\alpha _a-\beta _a)}\right] 
\end{eqnarray*}
and
\[ f_w\left( S_a(z)\right) =f_w(z)\ .\] 

Since we are summing over all the elements of the Schottky group, we then have
\begin{eqnarray*}
 & & f_w\left( T_a(z)\right) =T'_a(z)\left[ f_w(z)-\frac{\delta w_a}{\epsilon }\frac{(z-\alpha _a)(z-\beta _a)}{w_a(\alpha _a-\beta _a)}\right] \ ,\\ 
 & & f_w\left( S_a(z)\right) =f_w(z)\ ,
\end{eqnarray*}
which are the transformation properties we wanted. So we have obtained the Eichler integral that has the effect of changing infinitesimally the variables $w_a$ of a Riemann surface of genus $g$.

One way to verify this is a true Eichler integral is to take its third derivative. The result must be an automorphic form of weight $-2$. Taking the third derivative of (\ref{eq32}), we have
\begin{equation}
\label{eq33}
g_w(z)\equiv \frac{d^3}{dz^3}f_w(z) = \sum_b\frac{\delta w_b}{\epsilon w_b} \left[ \frac{(\alpha _b-\beta _b)}{(z-\alpha _b)(z-\beta _b)}\right] ^2\left[ \bar {\cal P}_{bc}\left( u(z)\right) -\bar {\cal P}''\left( u(z)\right) \right] \ ,
\end{equation}
where
\[ \bar {\cal P}_{bc}(u)=-\frac{\partial }{\partial u_c}\bar \zeta _b(u)\] 
is the hyperelliptic ${\cal P}$ function and
\[ \bar {\cal P}_{bc}''(u)=\frac{\partial ^2}{\partial u_c^2}\bar {\cal P}_{bc}(u_b)\ .\] 

It can be easily verified that $\bar {\cal P}_{bc}(u)$ and its derivatives are invariant under changes $z\rightarrow T_a(z)$ and $z\rightarrow S_a(z)$, so that the function $g(z)$ transforms like
\begin{equation}
\label{eq34}
g_w\left( T_a(z)\right) =\left( T'_a(z)\right) ^{-2}g_w(z)\ ,
\end{equation}
i.e. it is an automorphic form of weight $-2$. This confirms $f_w(z)$ as an Eichler integral.

\subsection{Example 6}

In the same way we found a function (Eichler integral) that changes the variables $w_a$ of a Riemann surface infinitesimally we now want to find a function that will change the variables $\alpha _a$ infinitesimally. This means this function must transform like
\[ f_\alpha \left( T_a(z)\right) =T'_a(z)f_\alpha (z)-\frac{\partial T_a(z)}{\partial \alpha _a}\frac{\delta \alpha _a}{\epsilon }\ ,\] 
for every $a=1,\dots ,g$, what implies that we want this function to transform like
\begin{equation}
\label{eq35}
f_\alpha \left( T_a(z)\right) =T'_a(z)\left[ f_\alpha (z)+\frac{(1-w_a)}{w_a(\alpha _a-\beta _a)^2}(z-\beta _a)^2\frac{\delta \alpha _a}{\epsilon }\right] \ .
\end{equation}
We also demand that
\[ f_\alpha \left( S_a(z)\right) =f_\alpha (z)\ .\] 
The method to obtain this function is analogous to the one used in the example 5, but now we use, instead of the Poincar\'e series $P_w(z)$, the series given in (\ref{eq5}):
\[ P_\alpha (z)=\sum_b \frac{\delta \alpha _b}{\epsilon }\frac{(1-w_b)}{w_b(\alpha _b-\beta _b)^2}(z-\beta _b)^2\ .\] 
The function (Eichler integral) with the properties we need is then given by
\begin{equation}
\label{eq37}
f_\alpha (z)=\sum_b \frac{\delta \alpha _b}{\epsilon }\frac{(1-w_b)}{w_b(\alpha _b-\beta _b)^2}(z-\beta _b)^2\bar \zeta _b(u)\ .
\end{equation}

\subsection{Example 7}

Similarly, if want a function that changes the variables $\beta _a$ infinitesimally, we then need it to transform like
\[ f_\beta \left( T_a(z)\right) =T'_a(z)f_\beta (z)-\frac{\partial T_a(z)}{\partial \beta _a}\frac{\delta \beta _a}{\epsilon }\ ,\] 
for every $a=1,\dots ,g$, i.e.
\begin{equation}
\label{eq38}
f_\beta \left( T_a(z)\right) =T'_a(z)\left[ f_\beta (z)+\frac{(1-w_a)}{w_a(\alpha _a-\beta _a)^2}(z-\alpha _a)^2\frac{\delta \beta _a}{\epsilon }\right] \ .
\end{equation}
We also demand that
\begin{equation}
\label{eq39}
f_\beta \left( S_a(z)\right) =f_\beta (z)\ .
\end{equation}
We then use the Poincar\'e series (\ref{eq6}),
\[ P_\beta (z)=\sum_b \frac{\delta \beta _b}{\epsilon }\frac{(1-w_b)}{w_b(\alpha _b-\beta _b)^2}(z-\alpha _b)^2\ ,\] 
and build the Eichler integral
\begin{equation}
\label{eq40}
f_\beta (z)=\sum_b \frac{\delta \beta _b}{\epsilon }\frac{(1-w_b)}{w_b(\alpha _b-\beta _b)^2}(z-\alpha _b)^2\bar \zeta _b(u)\ ,
\end{equation}
which has the correct transformation properties.

\vskip 0.5 cm 

\noindent {\large \bf Acknowledgements}

\vskip 0.3 cm 

The author would like to thank Professor P. C. West, under whose supervision this work was completed, and Dr. J. Harvey for directing me to Eichler integrals as possible solutions to my measure problem with strings.

I also would like to thank CAPES (Brazilian government) for financing this work. G.E.D.!

\appendix

\section{Some geometrical objects defined on a genus $g$ Riemann surface}

\subsection{Generalized $\theta $ functions}

The generalized $\vartheta $ function is defined in the following way \cite{8}:
\begin{eqnarray*}
\vartheta (v;\gamma ,\delta ) & = & \sum_{n=-\infty }^\infty \exp \left\{ \sum_{a,b=1}^g\left[ \frac{1}{2\pi i}v_a\eta _{(2)ab}v_b +\pi i(n_a+\delta _a)\tau _{ab}(n_b+\delta _b)\right] \right. \\ 
 & & \left. +\sum_{a=1}^g\left[ v_a(n_a+\delta _a)+2\pi i\gamma _a(n_a+\delta _a)\right] \right\} \ ,
\end{eqnarray*}
where the sum $\sum_{n=-\infty }^\infty $ means the sum over all $n_a$, $a=1,\dots ,g$. Here, $v_a$ and $v_b$ are the $g$ first Abelian integrals, $\eta _{(2)ab}$ is a symmetric $g\times g$ matrix, $n_a$ and $n_b$ are integers and $\gamma _a$ and $\delta _a$ are vectors with $g$ components that are the characteristics of the function.

If all the lements in the rows $p$ and $q$ are integers, this function transforms like \cite{8}
\[ \vartheta (v+\Omega ;\gamma ,\delta )=\exp \left\{ \sum_{a=1}^g\left[ H_a\left( v_a+\frac{1}{2}\Omega _a\right) -\pi ip_aq_a+2\pi i(q_a\delta _a-p_a\gamma _a)\right] \right\} \vartheta (v;\gamma ,\delta )\] 
where $v+\Omega $ stands for the sums $v_a+\Omega _a$ $(a=1,\dots ,g)$ where
\[ \Omega _a=2\pi i\sum_{b=1}^g\left[ 2\eta _{(1)ab}p_b+2\eta _{(2)ab}q_b\right] \] 
and
\[ H_a=\sum_{b=1}^g\left[ 2\eta _{(1)ab}p_b+2\eta _{(2)ab}q_b)\right] \] 
where $\eta _{(1)}$ and $\eta _{(2)}$ are $g\times g$ matrices and $p$, $q$ are $g$-vectors.

We can fix $\eta _{(2)ab}=0$. Thsi implies
\[ \eta _{(1)ab}=-\frac{1}{2}\delta _{ab}\ .\] 
We then have the following function:
\[ \theta (v;\gamma ,\delta ) = \sum_{n=-\infty }^\infty \exp \left\{ \sum_{a,b=1}^g \pi i(n_a+\delta _a)\tau _{ab}(n_b+\delta _b)+\sum_{a=1}^g\left[ v_a(n_a+\delta _a)+2\pi i\gamma _a(n_a+\delta _a)\right] \right\} \ ,\] 
which transforms like
\[ \theta (v+\Omega ;\gamma ,\delta )=\exp \left\{ \sum_{a=1}^g\left[ H_a\left( v_a+\frac{1}{2}\Omega _a\right) -\pi ip_aq_a+2\pi i(q_a\delta _a-p_a\gamma _a)\right] \right\} \theta (v;\gamma ,\delta )\ .\] 
This is the function that is generally reffered to as the generalized $\theta $ function in the literature. 

We may define the well-known object called the {\sl prime form} in terms of the generalized $\theta $ function, which is its most general definition. It is given by
\[ E(z,\zeta )=\theta \left( v(z)-v(\zeta )\right) \left[ \sum_{a=1}^g\partial _a\theta (0)w_a(z)\right] ^{-1/2}\left[ \sum_{b=1}^g\partial _b\theta (0)w_b(z)\right] ^{-1/2}\ ,\] 
where
\[ \partial _a\theta (0)\equiv \left. \frac{\partial }{\partial v_a}\theta (v)\right| _{v=0}\ \ ,\ \ \partial _b\theta (0)\equiv \left. \frac{\partial }{\partial v_b}\theta (v)\right| _{v=0}\ ,\] 
and $w_a(v)$, $w_b(v)$ are first Abelian differentials.

\subsection{Hyperelliptic $\zeta $ function}

We now define the {\sl hyperelliptic $\zeta $-function} \cite{8} \cite{9}:
\[ \zeta _a(v)=\frac{\partial }{\partial v_a}\ln \vartheta (v;\gamma ,\delta )\ .\] 
This function transforms in the following way:
\[ \zeta _a(v+\Omega ) = H_a+\zeta _a(v) = \sum_{b=1}^g \left[ 2\eta _{(1)ab}p_b +2\eta _{(2)ab}q_b\right] +\zeta _a(v)\ .\] 
The analogy with the one loop $\zeta $-functions is complete when we associate the matrices $\eta _{(1)}$ and $\eta _{(2)}$ with the numbers $\zeta (w_1/2)$ and $\zeta (w_2/2)$, respectively. We then have the identity
\[ \sum_{c=1}^g2\pi i\tau_{ac}\eta _{(2)cb}-2\pi i\eta _{(1)ab}=\delta _{ab}\ .\] 

A function that will be more useful to us is one that is invariant under a change $\Omega =2\pi iq_a$ so that we must fix $\eta _{(2)ab}=0$. Such a function, say $\bar \zeta _a$, is defined by
\[ \bar \zeta _a(v)=\frac{\partial }{\partial v_a}\ln \theta (v;\gamma ,\delta )\] 
and transforms like
\[ \bar \zeta _a(v+\Omega )=-p_a+\bar \zeta _a(v)\ .\] 
More particularly, this formula shows that the function $\bar \zeta $ is invariant under a change $\Omega _a=2\pi i\tau _{ab}p_b$ for $b\neq a$. It only changes, and then only by a constant term, under the transformation $\Omega _a=2\pi i\tau _{aa}p_a$ for any $p_a$.\vskip 2.0 cm

\subsection{Hyperelliptic ${\cal P}$ function}

We now define the hyperelliptic ${\cal P}$ function in the following way \cite{8} \cite{9}:
\[ {\cal P}_{ab}=\frac{\partial }{\partial v_b}\zeta _a(v)=\frac{\partial }{\partial v_b}\frac{\partial }{\partial v_a}\ln \theta (v)\ .\] 
This function is invariant under changes $\Omega _a=2\pi i\sum_{b=1}^g\tau _{ab}+2\pi iq_a$, i.e.
\[ {\cal P}_{ab}(v+\Omega )={\cal P}_{ab}\ .\]

\end{document}